\documentclass[pra,twocolumn,aps,amssymb,showpacs,superscriptaddress]{revtex4-1}

\usepackage{epsfig}
\usepackage{graphicx}
\usepackage{amsmath} 
\usepackage{amssymb}
\usepackage{enumerate}
\usepackage{hyperref}
\usepackage[normalem]{ulem}
\usepackage{cancel}

\usepackage{color}
\definecolor{rosso}{rgb}{1,0,0}
\definecolor{verde}{rgb}{0,1,0}
\definecolor{blue}{rgb}{0,0,1}
\definecolor{verdescuro}{rgb}{0,0.5,0.5}
\definecolor{rossoscuro}{rgb}{0.7,0.3,0}
\definecolor{bluscuro}{rgb}{0.3,0,0.7}
\definecolor{magenta}{rgb}{1,0,1}

\begin{document}

\title{Tunneling spectra of unconventional quasi-two-dimensional superconductors}

\author{L. Pisani}
\affiliation{Dipartimento di Fisica e Astronomia, Universit\`{a} di Bologna, 40127 Bologna, Italy}
\affiliation{INFN, Sezione di Bologna, 40127 Bologna, Italy}
\author{A. G. Moshe}
\affiliation{National Institute of Chemical Physics and Biophysics, 12618 Tallinn, Estonia}
\affiliation{Raymond and Beverly Sackler School of Physics and Astronomy, Tel Aviv University, Tel Aviv 6997801, Israel}
\author{P. Pieri}
\affiliation{Dipartimento di Fisica e Astronomia, Universit\`{a} di Bologna, 40127 Bologna, Italy}
\affiliation{INFN, Sezione di Bologna, 40127 Bologna, Italy}
\author{G. Calvanese Strinati}
\affiliation{School of Science and Technology, Physics Division, Universit\`{a} di Camerino, 62032 Camerino, Italy}
\affiliation{CNR-INO, Istituto Nazionale di Ottica, Sede di Firenze, 50125, Italy}
\author{G. Deutscher}
\affiliation{Raymond and Beverly Sackler School of Physics and Astronomy, Tel Aviv University, Tel Aviv 6997801, Israel}

\noindent

\begin{abstract}
Superfluid condensation can fundamentally be different from that predicted by the Bardeen-Cooper-Schrieffer (BCS) theory. 
In a broad class of low-carrier-density superconductors, such as granular aluminum (grAl), doped nitrides, and high-$T_{c}$ cuprates, tunneling experiments reveal strong rather 
than weak coupling,  as well as a conductance that does not return to that of the normal state upon approaching the critical temperature $T_{c}$. 
Here, we show that this behavior is in quantitative agreement with a tunneling theory that takes into account the large pairing-fluctuation effects that occur in the crossover region from weak coupling BCS 
to strong coupling Bose-Einstein condensation (BEC),
provided the coherence energy scale rather than the single-particle energy gap is used to evaluate the coupling ratio.
We also propose that the tendency towards strong coupling is a generic property of quasi-2D low-carrier density superconductors.
\end{abstract}

\maketitle

\emph{Introduction - \/} Lately, in condensed matter there has been an upsurge of interest in the BCS to BEC crossover owing to a growing experimental evidence for its occurrence in iron-based materials \cite{Kanigel-2012},
magic-angle twisted trilayer graphene \cite{Jarillo-Herrero-2021}, granular aluminum (grAl) \cite{Deutsher-2021}, and high-$T_{c}$ superconductors \cite{Harrison_Chan-2022,Deutscher-1999}.
For a recent review, see Ref.~\cite{Levin-RMP-2024}. 
These studies have followed the well-established achievements on the BCS-BEC crossover in ultra-cold Fermi gases obtained over the last twenty years \cite{Physics-Reports-2018}.

In particular, in Ref.~\cite{Harrison_Chan-2022} a number of thermodynamic measurements on high-$T_{c}$ cuprates have been organized in terms of the coupling ratio $2\Delta/k_{B}T_{c}$
 (where $\Delta$ is the low-temperature gap parameter and $k_{B}$ the Boltzmann constant), 
in such a way that the paired-fermion condensate was found to become optimally robust at the ``magic gap ratio''  $2\Delta/k_{B}T_{c} \approx 6$.
In that work, the gap parameter was taken as the single-particle excitation energy $\Delta_{\mathrm{p}}$ obtained, for instance, from tunneling experiments. 
We wish to point out here that this choice is by no means obvious because in the cuprates there is an additional energy scale that characterizes the condensed state. 
This is the coherence energy scale $\Delta_{\mathrm{c}}$ obtained, for instance, from Andreev-Saint-James point contact spectroscopy \cite{Deutscher-1999}. 
We argue that this is actually the correct energy scale to be used to evaluate the coupling ratio $2\Delta/k_{B}T_{c}$, because this parameter is meant to characterize the condensed state. 
We show that, if this choice is made, the tunneling conductance theory that we present is in good agreement with data from a variety of non conventional superconductors, including (besides the cuprates) granular aluminum, Li doped nitrides, and graphene, 
with values of the coupling ratio that do not exceed about $8$. 
This implies that known unconventional superconductors are essentially \emph{all\/} on the BCS side of the BCS-BEC crossover region, contrary to the conclusion reached in Ref.~\cite{Harrison_Chan-2022}.   
Examples about this important issue are given in Table~\ref{tab:values}, where the values of $2\Delta/k_{B}T_{c}$ for LSCO obtained when using for $\Delta$ either $\Delta_{\mathrm{c}}$ or $\Delta_{\mathrm{p}}$ are compared.
\begin{table}[h]

\centering
\begin{tabular}{|c|c|c|c|c|c|}
\hline
$x$ & $\Delta_{c}$ (meV) & $\Delta_{p}$ (meV) & $T_{c}$ (K) & $2\Delta_{c}/k_{B}T_{c}$ & $2\Delta_{p}/k_{B}T_{c}$ \\
\hline
0.08 & 4.0 & - & 9.6 & 9.6 & - \\
0.10 & 6.5 & - & 25.3 & 6.0 & - \\
0.12 & 6.0 & - & 26.0 & 5.4 & - \\
0.13 & 8.0 & 18.0 (*) & 29.2 & 6.4 & 14 \\
0.15 & 7.0 & 15.0 & 35.1 & 4.6 & 9.9 \\
0.2   & 6.5 & 7.0 & 27.8 & 5.4 & 5.8 \\
\hline
\end{tabular}
\caption{Values of $2\Delta/k_{B}T_{c}$, obtained when $\Delta$ is either $\Delta_{c}$ (from Andreev spectroscopy \cite{Gonnelli-2002})  
              or $\Delta_{p}$ (from tunneling experiments \cite{Gonnelli-2002}), for different $x$ values of La\textsubscript{2-x}Sr\textsubscript{x}CuO\textsubscript{4} samples.
              [(*) value extrapolated from Fig.~3 of Ref.~\cite{Gonnelli-2002}.]}             
\label{tab:values}
 
\end{table}
[Additional comparisons will be given in Fig.~\ref{Figure-4} below for the materials reported therein.]
The link between the coupling ratio $2\Delta/k_{B}T_{c}$ and the coupling parameter $(k_{F} a_{F})^{-1}$, used to describe the BCS to BEC crossover in ultra-cold Fermi gases and utilized in our numerical calculations, is discussed in Supplemental Material~\cite{footnote-SM}.

In this Letter, we present a number of tunneling spectra taken in grAl under a variety of physical conditions and organize them in terms of the coupling ratio $2\Delta/k_{B}T_{c}$ as above specified.
Our key finding is that the zero-bias conductance (ZBC) close to $T_{c}$ decreases as the coupling factor increases. 
This pseudo-gap effect is already clearly observed at moderate values of the coupling ratio. 
It is not to be confused with the large pseudo gap observed, for instance, in some under-doped cuprates.
In addition, we also report data extracted from the literature for the zero-bias conductance close to $T_{c}$ in $\mathrm{Bi2212}$ and graphene. 
All these transport measurements are shown to compare favorably with theoretical calculations for the differential conductance, that take into account the effects of pairing fluctuations throughout the BCS-BEC crossover,
provided the coherence energy scale $\Delta_{\mathrm{c}}$  is used to evaluate the coupling ratio.
Our suggestion, for a crossover between weak-coupling BCS to strong-coupling BEC revealed in grAl and other quasi-2D unconventional superconductors, thus rests on the basis of a quantitative comparison between experimental and theoretical tunneling conductances. 
This comparison represents the main contribution of this work.

From our analysis we conclude that in condensed-matter samples the crossover region between BCS and BEC regimes is  essentially exhausted for $2\Delta/k_{B}T_{c} \lesssim 8$.
We believe that the absence of condensed-matter samples with $2 \Delta/k_{B} T{c} \gtrsim 8$, which would be far deep in the BEC regime of tightly-bound pairs, is due to the charged nature of the fermions expected to pair up.
In this respect, we will also  point out that the quasi-2D nature of the materials here considered makes the values of $2 \Delta/k_{B}T_{c}$ somewhat larger than those expected for strictly 3D materials 
 \cite{footnote-SM}  (an issue that was not considered in Ref.~\cite{Harrison_Chan-2022}).
 
\emph{Comparison between experiment and theory - \/} 
The quantity to be calculated and compared with the experimental data is the differential conductance \cite{Tinkham-1980,Deutsher-2021}
\begin{equation}
\frac{d I}{d V}\!\left(eV\right) = B \int_{-\infty}^{+\infty} \!\!\!\! dE \, N_{s}(E) \left[ - \frac{\partial f_{F}(E + eV)}{\partial E} \right] ,
\label{differential-conductance}
\end{equation}
where $eV$ is the difference in the chemical potential across the superconducting-normal junction with an applied voltage $V$, $N_{s}(E)$ the quasi-particle density of states (DOS) at energy $E$,
$f_{F}(E) = 1/\left( e^{E/k_{B}T} + 1 \right)$ the Fermi function at temperature $T$, and $B$ a constant whose value we shall not be concerned with (and set equal to unity for simplicity).
In the following, we shall not distinguish between energy $E$ and frequency $\omega$ by setting $\hbar=1$.
The DOS function $N_{s}(\omega)$ in Eq.~(\ref{differential-conductance}) can, in turn, be expressed in terms of the \emph{single-particle spectral function\/} $A(\mathbf{k},\omega)$
by integrating it over the wave vector $\mathbf{k}$:
\begin{equation}
N_{s}(\omega) = \int \!\! \frac{d \mathbf{k}}{(2 \pi)^{3}} \, A(\mathbf{k},\omega) \, .
\label{DOS}
\end{equation}
Results for $A(\mathbf{k},\omega)$ in the superfluid phase, over the whole temperature range below $T_{c}$ and throughout the BCS-BEC crossover, were first obtained in Ref.~\cite{PPS-2004} in terms of 
a $t$-matrix approach, as an extension of an analogous approach in the normal phase above $T_{c}$ \cite{PPSC-2002}.
Recently, a refined version of this $t$-matrix approach was implemented to include the Gorkov-Melik-Barkhudarov (GMB) correction \cite{GMB-1961} across the BCS-BEC crossover, 
both in the normal \cite{PPPS-2018-I} and superfluid \cite{PPS-2018-II} phases.
This extended GMB approach has resulted in quite accurate comparisons with the experimental data for ultra-cold Fermi gases across the BCS-BEC crossover, both for the gap parameter $\Delta_{0}$ at low temperature \cite{Moritz-2022-I} 
and the critical temperature \cite{Koehl-2023}.
Here, we shall rely on this extended GMB approach to obtain the required single-particle spectral function for various couplings and temperatures \cite{footnote-analytic_continuation}.
Our theoretical analysis for a \emph{dynamical\/} quantity like the differential conductance (\ref{differential-conductance}) complements that utilized in Ref.~\cite{Harrison_Chan-2022} to compare with thermodynamic data, 
and is justified because for dynamical quantities the non-self-consistent $t$-matrix approach is expected to be more reliable than its fully self-consistent counterpart \cite{PPS-2019}.

\begin{figure}[t]
\begin{center}
\includegraphics[width=8.0cm,angle=0]{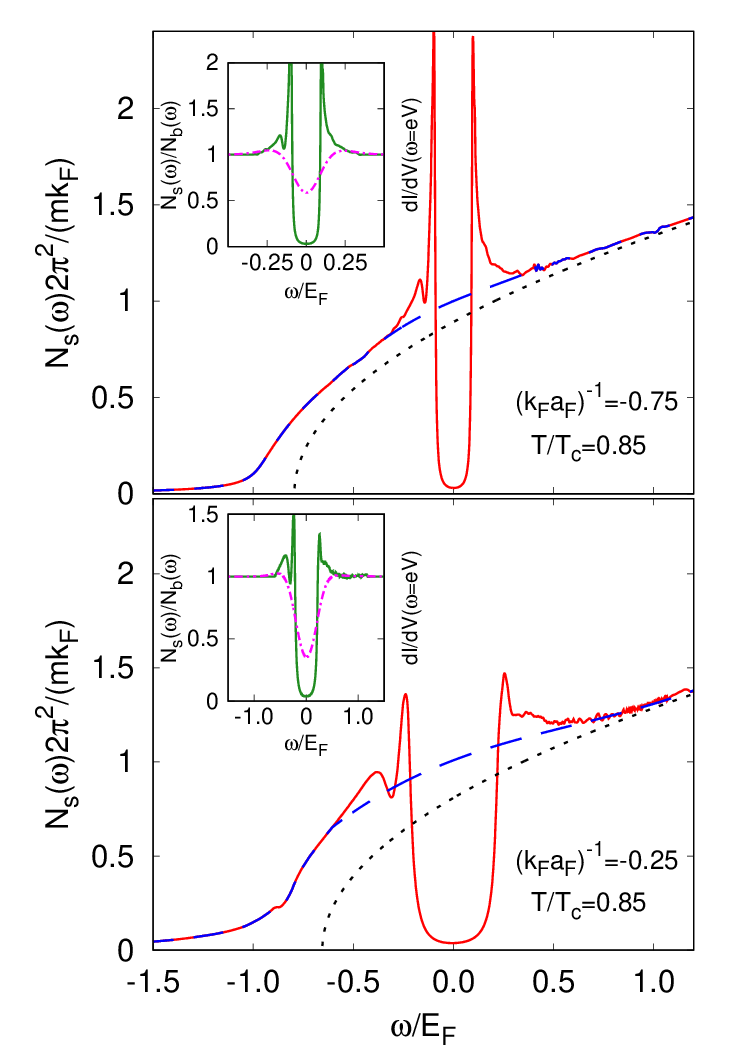}
\caption{DOS spectra $N_{s}(\omega)$ vs $\omega$ at $T = 0.85T_{c}$ for two different couplings in 3D (full lines).
              In both panels, the non-interacting form $N_{0}(\omega) = m^{3/2} \sqrt{2(\omega+\mu)}/(2 \pi^{2})$ is shown for comparison (dotted lines), which corresponds to a non-interacting system
              with the value of the chemical potential $\mu$ of the interacting system for given coupling.
              Plots are normalized by $N_{0}(\omega=0) = m k_{F}/(2 \pi^{2})$ where we have set $\mu = E_{F} = k_{F}^{2}/(2m)$. 
              Each panel also shows the associated background function $N_{b}(\omega)$ (dashed lines), obtained from the normalization procedure described in Ref.~\cite{footnote-SM}.
              The insets report the ratio $N_{s}(\omega)/N_{b}(\omega)$ (full lines) together with the corresponding normalized differential conductance 
              (dashed-dotted lines), obtained from Eq.~(\ref{differential-conductance}) with $B=1$ and $N_{s}(\omega)/N_{b}(\omega)$ replacing $N_{s}(\omega)$. [See Ref.~\cite{footnote-SM} for further details.]}
\label{Figure-1}
\end{center} 
\end{figure} 

Theoretical results obtained for the DOS spectra $N_{s}(\omega)$ in 3D are shown in Fig.~\ref{Figure-1}  at $T = 0.85 T_{c}$ for two couplings on the BCS side of unitarity.
In both cases, a dip is seen to occur in $N_{s}(\omega)$ about $\omega = 0$, accompanied by a two-peak structure  which is more visible for negative $\omega$.
Like in Ref.~\cite{PPS-2004}, we associate the dip to a finite value of the thermodynamic gap parameter, and the two-peak structure  to the presence of a pseudo-gap which develops upon approaching the normal phase.
Both features become progressively more pronounced for increasing coupling.
To highlight these features, in the insets of Fig.~\ref{Figure-1}  the DOS spectra are normalized by the background function $N_{b}(\omega)$ described in Ref.~\cite{footnote-SM},
which aims at eliminating any irrelevant reference to the 3D background of the theoretical DOS spectra. 
This is achieved by taking $N_b(\omega)$  to coincide with $N_s(\omega)$ at large frequencies, where kinetic energy dominates over interaction,  with a simple interpolation at low frequencies.
This procedure enables us to extract in an effective way the quasi-2D character of the spectra, with the pseudo-gap associated with pairing fluctuations being regarded as a local quantity that should not appreciably differ 
in 3D and quasi-2D \cite{Moritz-2022-III} (cf. also the discussion below and Ref.~\cite{footnote-SM}).
Further results for the temperature dependence of the DOS spectra at fixed coupling are reported in Ref.~\cite{footnote-SM}, where the persistence of the pseudo-gap upon approaching $T_{c}$ is evident.

\begin{figure}[t]
\begin{center}
\includegraphics[width=8.1cm,angle=0]{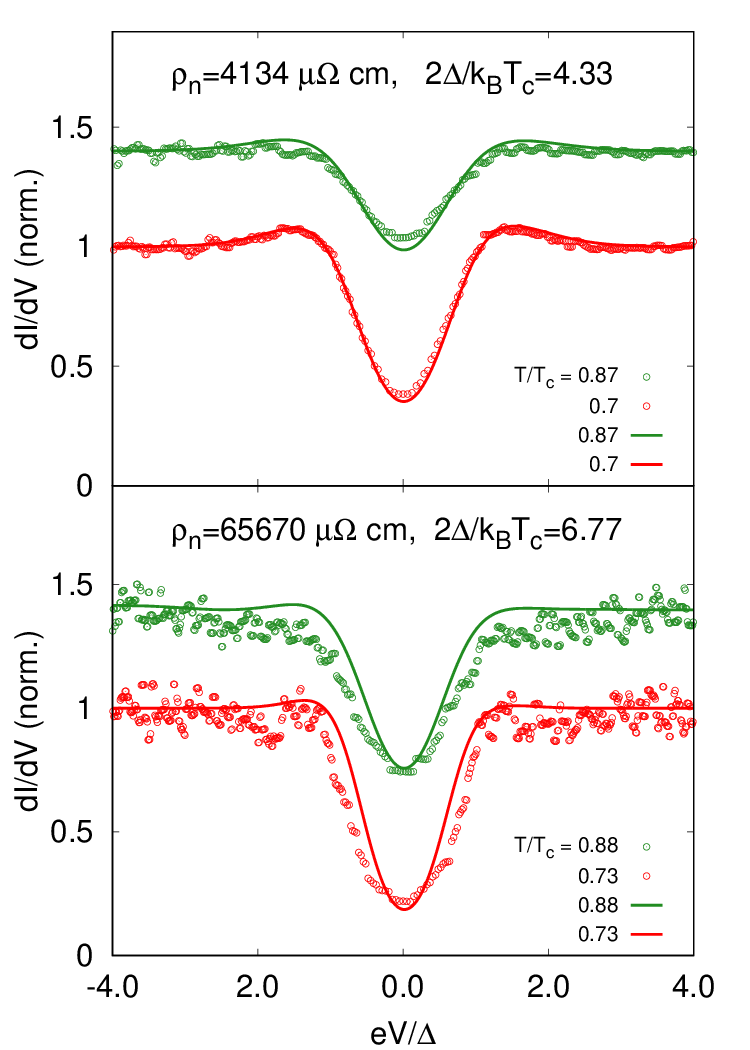}
\caption{Normalized differential conductance vs voltage for two characteristic measured samples, identified by their resistivity $\rho_{\mathrm{n}}$ and coupling ratio $2\Delta/k_{B}T_{c}$.
              Open circles are experimental data and full lines are theoretical results calculated for the values $(k_{F}a_{F})^{-1} = (-0.75,-0.25)$ of the nominal coupling (from top to bottom). 
              Experimental and theoretical data are compared with each other by normalizing the frequency by the respective values of the gap parameter $\Delta_{0}$ at low temperature.
              Results for the larger temperatures are shifted upwards for clarity.
              [See Ref.~\cite{footnote-SM} for further details.]}
\label{Figure-2}
\end{center} 
\end{figure}

To draw a meaningful comparison between experiment and theory, the experimental differential conductance $dI/dV$ spectra have been normalized by fitting the data in the range of $1.5<\left|V\right|<3.0$ mV 
to the form $dI/dV\propto\left|V\right|$ and then dividing the measured $dI/dV$ by the extrapolated values obtained from the fit for $\left|V\right|<3.0$ mV. 
The experimental differential conductances below $T_c$ for two representative samples are shown in Fig.~\ref{Figure-2}  at two selected temperatures,
where they are plotted as a function of bias normalized by their respective zero-temperature gap $\Delta$. 
The lower resistivity sample (with $T_{c} =  2.31K$ and $2\Delta/k_{B} T_{c} = 4.33$) appears to be close to the BCS side (upper panel), 
while the higher resistivity sample (with $T_{c }= 1.37K$ and $2\Delta/k_{B} T_{c} = 6.77$) to be close to the unitarity regime (lower panel), with the estimated value $\Delta/E_{\rm F} \simeq 0.35$ \cite{PPS-2018-II,Moshe-2019}.  
In all cases, the experimental data are seen to closely match the overall theoretical trend, signaling further that the behavior of both considered samples deviates from what would be expected in 
the BCS weak-coupling limit.The procedure for associating to a given sample the nominal value of the coupling $(k_{F}a_{F})^{-1}$ is discussed in Ref.~\cite{footnote-SM}, 
where the quasi-2D character of the samples is taken into account in an effective way. 
Note how, for the differential conductance,  the convolution with the derivative of the Fermi function in Eq.~(\ref{differential-conductance}) smears out the two-peak structure of Fig.~\ref{Figure-1} for $N_{\rm s}(\omega)$.
Note further that,  by comparing tunnel and optical gaps, in Ref.~\cite{Deutsher-2021} we concluded that atomic disorder effects are negligible in grAl. This excludes disorder effects \cite{Altshuler-1980} as an explanation for the pseudo-gap observed in grAl. 
\begin{figure}[t]
\begin{center}
\includegraphics[width=9.0cm,angle=0]{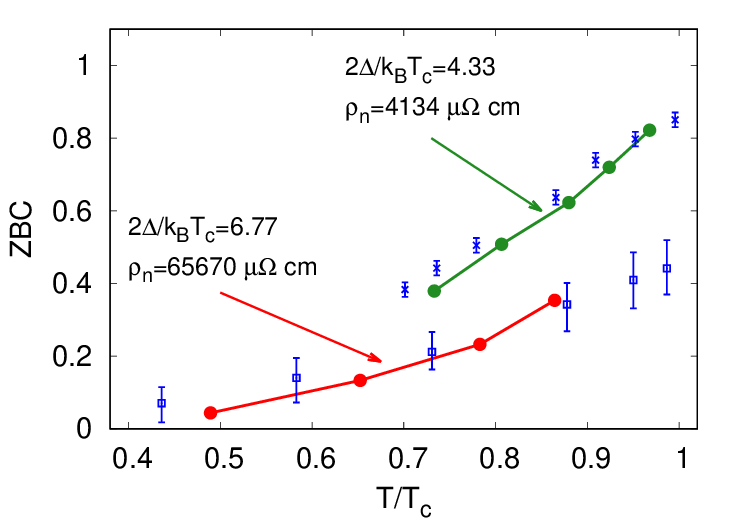}
\caption{Zero-bias conductance (ZBC) vs reduced temperature for the same samples of Fig.~\ref{Figure-2}. 
              Crosses and empty squares with error bars are experimental data, and filled circles are theoretical results obtained for the same values of the nominal coupling $(k_{F} a_{F})^{-1}$ of Fig.~\ref{Figure-2}.
              In all cases, the temperature is in units of the respective value of the critical temperature $T_{c}$.}
\label{Figure-3}
\end{center} 
\end{figure}

A further measurable quantity of interest is the zero-bias conductance (ZBC), obtained as the limit $dI/dV|_{V \rightarrow 0}$. 
Figure~\ref{Figure-3}  shows the dependence of ZBC on temperature for the same samples of Fig.~\ref{Figure-2}.
(For the higher $\rho_{\mathrm{n}}$, the ZBC value extrapolated to $T=0$ from the measured ones has been subtracted off to avoid spurious (leakage) effects at the contacts. Additional spurious effects, possibly arising from the formation of  pinholes in the Al oxide barrier layer, are minimized by our oxidization procedure \cite{Deutsher-2021}.)
The corresponding theoretical dependence, calculated for the same nominal couplings of Fig.~\ref{Figure-2}, is seen to follow the same trend as the experiment.
For the higher $\rho_{\mathrm{n}}$, note the strong depletion of ZBC for both experiment and theory, which extrapolates to a value less than $0.5$ upon approaching $T_{c}$.
For the lower $\rho_{\mathrm{n}}$, this extrapolation gets instead close to unity, a value which is expected to be reached in the BCS limit of the BCS-BEC crossover. 

\begin{figure}[t]
\begin{center}
\includegraphics[width=8.8cm,angle=0]{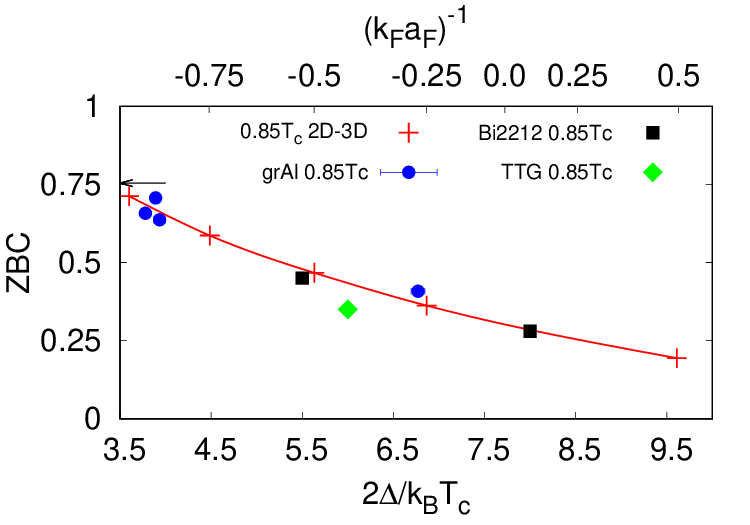}
\caption{ZBC vs coupling ratio.
              The dependence of the zero-bias conductance (ZBC) on the coupling ratio $2 \Delta/k_{B}T_{c}$ for grAl at $0.85 T_{c}$ (filled circles)
              is compared with the theoretical results obtained by the extended GMB approach also at $0.85 T_{c}$ (crosses and full line).
              Further experimental values are shown for Bi2212 (filled squares) [from Ref.~\cite{Krasnov-2009}] and for twisted trilayer graphene (filled diamond) [from Ref.~\cite{Kim-2022}], both at $0.85 T_{c}$.
              The arrow signals the BCS mean-field result at $0.85 T_{c}$ for coupling $(k_{F} a_{F})^{-1}=-1.5$.
              The upper horizontal axis reports the values of the nominal coupling $(k_{F}a_{F})^{-1}$ associated with the values of the coupling ratio $2 \Delta/k_{B}T_{c}$ according to the procedure summarized in
              Fig.~9 of Ref.~\cite{footnote-SM}.}
\label{Figure-4}
\end{center} 
\end{figure} 

Finally, Fig.~\ref{Figure-4}  shows the experimental values for ZBC versus the coupling ratio $2\Delta/k_{B}T_{c}$ for several samples at $T = 0.85 T_{c}$,
together with the corresponding theoretical results also at $T = 0.85 T_{c}$. 
In this way, we avoid entering the critical region close to $T_{c}$, where the theory would need to be refined by including additional fluctuation corrections \cite{footnote-SM}.
This critical region is larger at unitarity than in the BCS regime \cite{Taylor-2009}, as implicitly confirmed by the different extensions toward $T_{c}$ of the theoretical results reported in Fig.~\ref{Figure-3}  for the two samples.
One sees from Fig.~\ref{Figure-4}  that the ZBC decreases as the coupling ratio increases in good agreement with the theoretical trend, being reduced by about half when reaching $2 \Delta/k_{B}T_{c} \simeq 6$.
Figure~\ref{Figure-4}  also shows that the above overall trend is not unique to grAl by considering ZBC values for two other compounds, Bi2212 \cite{Krasnov-2009} 
and twisted trilayer graphene (TTG) \cite{Kim-2022}.
For both of them the tunneling gap is substantially larger than the coherence energy scale obtained by point contact spectroscopy, such that for TTG $2\Delta_{\mathrm{c}}/k_{B}T_{c} = 6$ while $2\Delta_{\mathrm{p}}/k_{B}T_{c} = 15-19$,
for under-doped Bi2212 $2\Delta_{\mathrm{c}}/k_{B}T_{c} = 8$ while $2\Delta_{\mathrm{p}}/k_{B}T_{c} = 10$,
and for over-doped Bi2212 $2\Delta_{\mathrm{c}}/k_{B}T_{c} = 5.5$ while $2\Delta_{\mathrm{p}}/k_{B}T_{c} = 8.7$.
Good agreement with theory is only obtained if the $\Delta_{\mathrm{c}}$ is used to evaluate the coupling ratio. 
On the other hand, Table I suggests that for our grAl samples, for which the coupling ratio as measured by tunneling does not exceed 7,  differences between $\Delta_{\mathrm{p}}$ and $\Delta_{\mathrm{c}}$  should be minor. 
We further note that, besides Bi2212 and TTG, grAl is also a quasi-2D superconductor \cite{Deutscher-1977}.
In this respect, one should mention that, away from the weak-coupling (BCS) regime, the values of the coupling ratio $2 \Delta/k_{B}T_{c}$ change appreciably when calculated in 3D or in the strictly-2D case of a BKT transition.
This change is here taken into account when associating the experimental value of $2\Delta/k_{B}T_{c}$ with the nominal coupling $(k_{F}a_{F})^{-1}$ used for theoretical calculations \cite{footnote-SM}.

We note that, when the proper energy scale is used to evaluate the coupling ratio, this ratio never exceeds the value of about $8$ which corresponds to the unitary limit, 
 suggesting that this is a universal property of quasi-2D superconductors close to a Mott transition, triggered by Coulomb interaction.

\emph{Discussion and outlook - \/}
The extensive comparison we have presented of the experimental and theoretical spectra for the differential conductance is made between experimental samples
(like grAl films) that are quasi-2D systems and theoretical calculations that are performed for a 3D system.
However, there is intrinsically no contradiction in making this comparison, inasmuch as the experiment seems to have access to a pseudo-gap phase related to pairing fluctuations 
which survive across $T_{c}$. 
On physical grounds, what should, in fact, be important is the \emph{local\/} behavior of pairing correlations occurring inside a Cooper pair (or, at most, among nearby Cooper pairs).
Under these circumstances, pseudo-gap phenomena occurring in a fermionic superfluid should be envisaged as due to persistency of ``local-pairing order” across $T_{c}$, 
that survives even when (off-diagonal) long-range order is lost. 
Accordingly, one expects that there should not be too much of a difference between 3D and quasi-2D calculations for this local property.
In this context, one should mention a recent experimental work \cite{Moritz-2022-III} where it was shown that dimensionality has only limited influence on the stability of strongly interacting fermionic superfluids,
in the sense that there is nothing significantly new in the superfluid properties of 3D and 2D Fermi superfluids.

These arguments are reinforced by the theoretical work of Ref.~\cite{Witkowski-2020} where the effect of a weak inter-layer coupling on BE condensates was studied,
with the result that a small inter-layer coupling was found sufficient to preserve BEC with an only slightly reduced condensation temperature; as well as by the
analogy with a layered anti-ferromagnet mentioned in Ref.~\cite{footnote-SM}, where local anti-ferromagnetic correlations surviving above the N\'{e}el temperature 
were also seen not to markedly depend on the ratio between the intra- and inter-layer couplings.
In all cases, local correlations do eventually allow to achieve symmetry breaking at a slightly reduced temperature.

Our results corroborate the mounting experimental evidence that a combination of small carrier density and reduced dimensionality acts to favor superconductivity \cite{Qiu-2021}.
This is the case for systems as diverse as the high-$T_{c}$ cuprates, grAl, and twisted trilayer graphene.
Nitrides such as Li\textsubscript{x}HfNCl were not considered in Fig.~\ref{Figure-4}  
because their coherence energy scale has not been reported so far, but the same combination holds for them too.
One is thus led to look for a general (yet simple) argument to explain these observations. 
This argument may be provided by the jellium model for superconductivity \cite{Nozieres-Pines-1958,DeGennes-1999,Deutscher-2006}. 
In this model, the electron-electron attraction $V$ increases as the carrier density decreases. 
In 3D this increase is counterbalanced by a reduced density of states, so that the BCS parameter $N_{0}V$ and the critical temperature get eventually reduced for decreasing carrier density. 
This is not the case in 2D when the density of states remains constant, such that both $N_{0}V$ and $V$ should increase for reduced carrier density. 
As a consequence, the interaction parameter $V$ might be strong enough to promote a crossover from BCS to BEC.
There are, however, a few problems with this argument. 
In 2D localization sets in even for extremely small disorder and might prevent superfluidity. 
In addition, BEC does not occur in strictly 2D at finite temperature. 
These difficulties are in principle removed by the quasi-2D approach.

\emph{Conclusions - \/}
The detailed experimental tunneling results on quasi-2D grAl, Bi2212, and twisted trilayer graphene presented in this work appear to exhibit the main features expected for a BCS-BEC crossover.
Namely, a coupling ratio $2\Delta/k_{B}T_{c}$ that evolves continuously from the weak-coupling (BCS) limit value to larger values typical of the crossover regime, and at the same time monotonically decreasing values 
of Zero Bias Conductance close to $T_{c}$ with increasing coupling ratio. 
We have further shown that, for three families of small-carrier density quasi-2D systems here considered, there appears to be a universal correlation between the coupling ratio and the Zero Bias Conductance, 
and that this correlation is in agreement with theoretical results obtained for the BCS-BEC crossover provided the coherence energy scale is used to evaluate the coupling ratio.

\emph{Acknowledgements - \/} 
A.~G.~M. acknowledges support by the Estonian Research Council under Grant No. MOBJD1103.


\clearpage

\begin{center}
{\bf Supplemental Material}
\end{center}

\section{Introduction}
\label{sec:introduction}

This Supplemental Material contains the following: 
Sec.~\ref{sec:experimental_framework} provides experimental details on the sample preparation;
Sec.~\ref{sec:coupling-parameter_vs_coupling-ratio} addressed the relation between the coupling parameter $(k_{F} a_{F})^{-1}$ utilized for ultra-cold Fermi gases and the coupling ratio $2 \Delta/k_{B}T_{c}$ adopted for condensed-matter superconductors;
Sec.~\ref{sec:theoretical_framework} discusses the normalization procedure adopted in the theoretical DOS spectra to allow for a direct comparison with the experimental spectra of the differential conductance;
Sec.~\ref{sec:critical_region} discusses the emergence of a critical region upon approaching the critical temperature $T_{c}$, where the theory would need to be refined by including further fluctuation corrections beyond those
explicitly considered;
Sec.~\ref{sec:2D:contact_potential} discusses the issue of a meaningful comparison between the 3D and 2D values of the coupling ratio $2\Delta/k_{B}T_{c}$, and introduces an ``effective'' intermediate dimensionality;
Sec.~\ref{sec:anti_ferromagnet} makes an analogy with the dimensional crossover occurring in a layered anti-ferromagnet, for which similar results are obtained for local properties calculated between 3D and quasi-2D;
Sec.~\ref{sec:coherence_length} points out that the tendency toward a minimum in the coupling-dependence of the low-temperature coherence length is a fingerprint of the approach to the unitarity regime. 

\vspace{-0.5cm}
\section{Experimental details}
\label{sec:experimental_framework}

The grAl films were prepared by thermally evaporating clean Al pellets under controlled O\textsubscript{2} pressure onto liquid nitrogen cooled substrate, as described in previous work \cite{MoshePRB-2021s}. 
By varying the O\textsubscript{2} pressure, we obtain wide range of samples, showing a dome-shaped phase diagram of the critical temperature T\textsubscript{c} vs the normal state resistivity $\rho_{\mathrm{n}}$. 
The tunneling junctions were prepared by evaporating a strip of 50 nm of grAl, letting it oxidize in ambient conditions and then evaporating two clean Al strips in perpendicular to the grAl strip. 
The current ($I$) vs voltage ($V$) and $dI/dV$ vs $V$ were measured at temperatures close to and below T\textsubscript{c} in a \textsuperscript{4}He bath cryostat or a commercial Quantum
Design \textsuperscript{3}He probe. 
More details can be found in previous work \cite{MoshePRB-2021s}.

\section{Coupling parameter vs coupling ratio}
\label{sec:coupling-parameter_vs_coupling-ratio}

For 3D ultra-cold Fermi gases, it is common practice (both theoretically and experimentally) to identify the interaction part of the Hamiltonian in terms of the coupling parameter $(k_{F} a_{F})^{-1}$ \cite{Physics-Reports-2018s}.
Here, $k_F=(3\pi^2 n)^{1/3}$ is the Fermi wave vector with number density $n$ and $a_{F}$ the scattering length of the fermionic two-body problem in vacuum. 
In 3D ultra-cold Fermi gases, the coupling parameter $(k_{F} a_{F})^{-1}$ 
ranges from $(k_{F}\, a_{F})^{-1} \lesssim -1$ in the BCS regime when $a_{F} < 0$, to $(k_{F}\, a_{F})^{-1} \gtrsim +1$ in the BEC regime when $a_{F} > 0$, across unitarity when $|a_{F}|$ diverges \cite{Physics-Reports-2018s}.

In condensed matter, whereby the parameter $(k_{F} a_{F})^{-1}$ is not directly accessible, one prefers instead to use the coupling ratio $2 \Delta/k_{B}T_{c}$ which equals about $3.5$ for weak-coupling (BCS) superconductors 
but is expected to increase for stronger couplings.
Here and in the following, $\Delta$ refers to the low-temperature gap parameter 
associated with the coherence energy scale \cite{Deutsher-2019s} (as discussed in detail in the Main Text) 
and $T_{c}$ is the superfluid critical temperature ($k_{B}$ being the Boltzmann constant).
In unconventional superconductors such as grAl \cite{Deutsher-2019s} and the high-$T_{c}$ cuprates \cite{Harrison_Chan-2022s}, large values of the ratio $2 \Delta/k_{B}T_{c}$ have been reported 
(in particular, for grAl this ratio was seen to increase monotonically as the metal to insulator transition is approached). 
A correspondence between the coupling parameter $(k_{F} a_{F})^{-1}$ and the coupling ratio $2 \Delta/k_{B}T_{c}$ was pointed out in Ref.~\cite{Deutsher-2019s} (cf. Fig.~9 therein), 
where the value $2 \Delta/k_{B}T_{c} \simeq 6.5$ was associated with the unitarity point $(k_{F}\, a_{F})^{-1}=0$ in 3D.

In Sec.~\ref{sec:2D:contact_potential} we will extend the correspondence between the coupling ratio $2 \Delta/k_{B}T_{c}$ and the coupling parameter for ultra-cold Fermi gases to the 2D case, 
and further introduce an ``intermediate” dimensionality between 3D and 2D which appears relevant to the condensed-matter samples analyzed in the Main Text.

\vspace{-0.5cm}
\section{Normalization procedure of \\ the DOS spectra}
\label{sec:theoretical_framework}

We are interested in the pseudo-gap-like features associated with a dip about $\omega = 0$ and a ``two-peak structure'' for negative $\omega$, which are present in the DOS spectra $N_{s}(\omega)$ up to the critical temperature $T_{c}$.
Accordingly, we need to isolate these features from the ``outer'' background, which extends in the outer spectral region and is of no specific physical interest.

To this end, we introduce a suitable \emph{normalization procedure} of the DOS spectra, in analogy to the extrapolation of the background (normal-state) conductance required to extract relevant information from asymmetric spectra of high-temperature superconductors \cite{Fischer-1995s,Fischer-2007s,TTG-2022s}.
In the present case, this normalization procedure is needed to produce a meaningful comparison between experiment and theory on the differential tunneling conductance, owing to the different dimensional and geometrical constraints of the theoretical (3D) modeling with respect to the actual (quasi-2D) materials.
In particular, the aim here is to eliminate any irrelevant reference to the 3D background of the theoretical DOS spectra, as it is done in Fig.~1  of the Main Text.
This makes the normalized spectra to resemble as much as possible the quasi-2D case, with a ``flat'' background that shows up in the experimental spectra for the differential conductance.
Yet, this normalization is not expected to influence in a relevant way the central part of interest of the spectra about $\omega \approx 0$, where pseudo-gap-like features are expected to show up.
From a physical point of view, we expect this central part of the spectra should not be too much affected by the spatial environment being 3D or quasi-2D, as discussed in Secs.~\ref{sec:2D:contact_potential} and \ref{sec:anti_ferromagnet} below.

In practice, we set up the \emph{normalization procedure\/} of the DOS spectra $N_{s}(\omega)$ for given coupling and temperature according to the following steps:

\noindent
(i) Determine two frequencies $\omega_{c}^{\pm}$, such that
$\omega_{c}^{-}$ ($\omega_{c}^{+}$) corresponds to the upper (lower) limit of the range $\omega < \omega_{c}^{-}$ ($\omega_{c}^{+} < \omega$) where $N_{s}(\omega)$ does not appear to depend on temperature.

\noindent
(ii) Introduce a ``breach'' in the spectrum of $N_{s}(\omega)$ for $\omega_{c}^{-} \le \omega \le \omega_{c}^{+}$, and use a cubic spline interpolation over the wider frequency interval $- 3 E_{F} \le \omega \le + 3 E_{F}$ that includes this breach.
This is done to determine the background function $N_{b}(\omega)$, which approximates $N_s(\omega)$ for all $\omega$ outside the breach and interpolates inside the breach region.

\noindent
(iii) Normalize the DOS spectrum $N_{s}(\omega)$ for \emph{all\/} $\omega$, by dividing it by the background function $N_{b}(\omega)$ determined in step (ii).

\noindent
(iv) Obtain the corresponding ``normalized'' differential conductance by replacing $N_{s}(\omega) \rightarrow N_{s}(\omega)/N_{b}(\omega)$ in Eq.~(1) of the Main Text.

\begin{figure}[t]
\begin{center}
\includegraphics[width=7.2cm,angle=0]{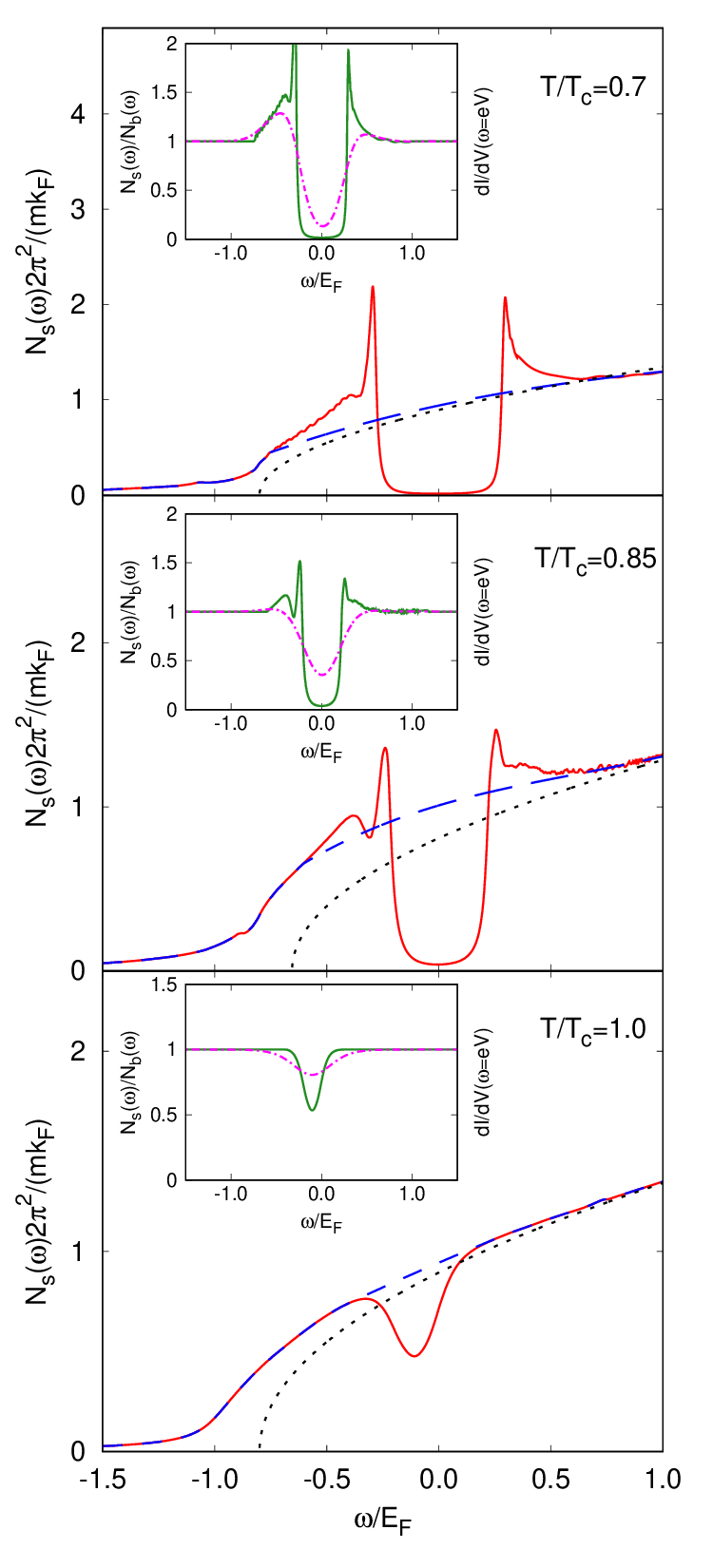}
\caption{The quantities, that were shown in Fig.~1 of the Main Text at $T_{c}$ for two couplings, are here shown for the coupling $(k_{F} a_{F})^{-1}=-0.25$ and three temperatures $T/T_{c} = (0.7,0.85,1.0)$.
              Conventions for the various quantities are the same as those of Fig.~1 of the Main Text.}
\label{Figure-1-SM}
\end{center} 
\end{figure} 

As an example of how this normalization procedure works in practice, Fig.~1 of the Main Text shows the background functions $N_{b}(\omega)$ (dashed lines) that correspond to the bare DOS spectra $N_{s}(\omega)$ obtained at $T = 0.85 T_{c}$ (full lines) for two relevant couplings.
These plots make evident the way how the background function $N_{b}(\omega)$ acts to avoid the breach, which is artificially introduced in the bare $N_{s}(\omega)$ spectra over the interval 
$\omega_{c}^{-} \le \omega \le \omega_{c}^{+}$.
In addition, in each panel of Fig.~1 of the Main Text the inset reports the ratio $N_{s}(\omega)/N_{b}(\omega)$ (full lines) and the corresponding normalized differential conductance, which is obtained from Eq.~(1) of the Main Text with the ratio $N_{s}(\omega)/N_{b}(\omega)$ replacing $N_{s}(\omega)$ (dashed-dotted lines).
Note also that, by this procedure, the tunneling suppression occurring in the spectra of the normalized differential conductance shown in the insets of Fig.~1 of the Main Text naturally evolves from V-like-shaped to U-like-shaped as the coupling moves from the BCS side toward unitarity, in line with a recent observation made in twisted trilayer graphene for similar spectra \cite{TTG-2022s}.

This procedure can be repeated for each coupling over the whole temperature range $0 \le T \le T_{c}$.
As an example, Fig.~\ref{Figure-1-SM} shows plots analogous to those of Fig.~1 of the Main Text, which are now obtained for the coupling $(k_{F} a_{F})^{-1}=-0.25$ at three different temperatures.
Note how the suppression of weight about $\omega = 0$ and the ``two-peak structure"  for negative $\omega$ evolve in temperature upon approaching $T_{c}$.
The insets make evident the corresponding temperature evolution of the spectra for the normalized differential conductance.
Note how the convolution made in Eq.~(1) of the Main Text smoothens out the small noisy fluctuations present in the density of states $N_{s}(\omega)$, which stem from Eq.~(2) of the Main Text when the spectral function $A(\mathbf{k},\omega)$ 
presents sharp features at low temperature in the superfluid phase.   
These are the theoretical frequency spectra that in the Main Text are compared with the experimental data for the differential conductance as a function of the bias voltage.

\begin{figure}[t]
\begin{center}
\includegraphics[width=8.5cm,angle=0]{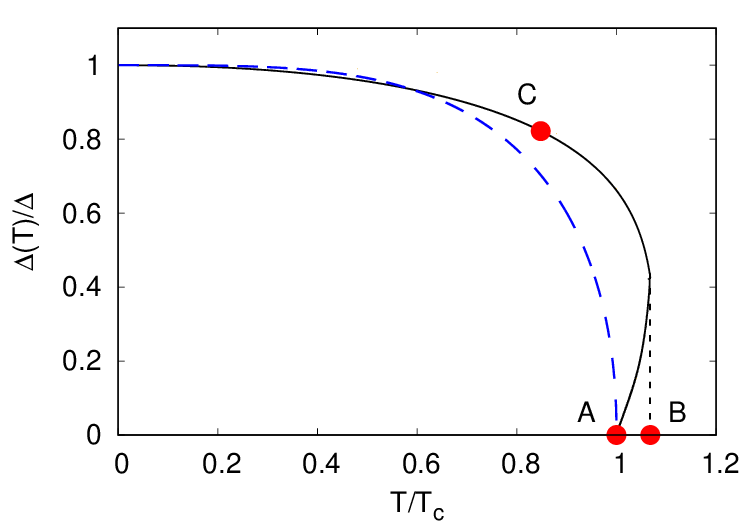}
\caption{The temperature dependence $\Delta(T)$ of the gap parameter (normalized to its zero-temperature value $\Delta$) is shown to results in a re-entrant behavior close to $T_{c}$ 
              for coupling $(k_{F} a_{F})^{-1} = -0.25$ (full line).
              The meaning of the three points $A$, $B$, and $C$ (filled circles) which bound this re-entrant region is explained in the text.
              The vertical dotted line shows how the point $B$ is identified from the re-entrant behavior.
              Also shown is the corresponding profile of $\Delta(T)$ for coupling $(k_{F} a_{F})^{-1} = -0.75$ closer to the BCS regime (dashed line), for which the re-entrant behavior is essentially absent.}
\label{Figure-2-SM}
\end{center} 
\end{figure} 

\section{Emergence of a critical region upon approaching $T_{c}$}
\label{sec:critical_region}

The theoretical results presented in the Main Text are based on the $t$-matrix approach for the superfluid phase of Ref.~\cite{PPS-2004s}, once updated to the extended Gorkov-Melik-Barkhudarov (GMB) approach of Refs.~\cite{PPPS-2018-Is,PPS-2018-IIs}, 
whereby pairing fluctuations are included in the diagrammatic theory over and above the standard mean-field approach \cite{Mahan-2000s}.
Quite generally, the inclusion of pairing fluctuations affects the expression of the differential conductance, either directly at the single-particle level resulting into the DOS contribution that is explicitly considered in Eq.~(1) of the Main Text, 
or at the two-particle level in what are referred to as the Aslamazov-Larkin (AL) and Maki-Thompson (MT) diagrammatic contributions \cite{LV-2009s}.

The AL and MT contributions, however, were not considered in the Main Text for the calculation of the differential conductance.
As a matter of fact, in the literature the AL and MT contributions have thus far been explicitly calculated mostly in the normal phase above $T_{c}$ and in the weak-coupling (BCS) limit only, where the presence of an underlying Fermi surface introduces
key simplifications which make analytic calculations feasible \cite{Varlamov-1993s}.
Extending these calculations to the superfluid phase below $T_{c}$, and in addition throughout the BCS-BEC crossover, would thus require us to undergo major analytic and numeric efforts which by far exceed the purposes of the present work.
Nevertheless, what one learns from the available calculations of the AL and MT contributions in the normal phase above $T_{c}$, is that they become as important as the DOS contribution only when entering the critical region close to $T_{c}$
such that the interplay among the DOS, AL, and MT contributions becomes crucial in the critical region, while outside this region the DOS contribution dominates \cite{LV-2009s}.
By expecting on physical grounds these arguments to be transposed to the superfluid phase below $T_{c}$, this is one reason why in the Main Text we have decided to avoid entering the critical region when approaching $T_{c}$ from below,
thus limiting the theoretical calculations in practice to $T \lesssim 0.85 T_{c}$.

But there is an additional reason why the critical region should prudently be avoided (especially near unitarity) even when keeping the DOS contribution only, as we did in the Main Text.
This is related to a peculiar temperature behavior of the gap parameter $\Delta(T)$ for $T$ close to $T_{c}$ upon approaching unitarity, which results from theoretical calculations based on the $t$-matrix approximation irrespective of whether its non-self-consistent \cite{PPS-2004s} or fully-self-consistent \cite{Haussmann-2007s} version is utilized, and even when the GMB correction is included over the whole BCS-BEC crossover 
\cite{PPPS-2018-Is,PPS-2018-IIs}.
A typical example of this behavior is shown in Fig.~\ref{Figure-2-SM}  for the function $\Delta(T)$ obtained at coupling $(k_{F} a_{F})^{-1} = - 0.25$ within the extended GMB approach of Ref.~\cite{PPS-2018-IIs}.
It is evident from this plot that, close to $T_{c}$, the gap parameter is a multivalued function of temperature, with a re-entrant behavior reminiscent of a first-order transition.
Since this feature is shared by all versions of the $t$-matrix approach as already mentioned, its healing should possibly require one to include additional classes of diagrammatic contributions at the single-particle level, 
beyond the ladder structure of the $t$-matrix or its variants.
Note also from Fig.~\ref{Figure-2-SM}  that the re-entrant behavior in $\Delta(T)$ is essentially absent for coupling $(k_{F} a_{F})^{-1} = - 0.75$ closer to the BCS regime.
These features are consistent with the finding that the critical region is larger at unitary than in the BCS regime \cite{Taylor-2009s}.

The presence of the re-entrant behavior has led Ref.~\cite{Haussmann-2007s} to identify two critical temperatures: 
(i) A ``lower'' $T_{c}^{-}$, corresponding to the point $A$ in Fig.~\ref{Figure-2-SM}  and calculated either by implementing the Thouless criterion from the normal phase or by following the profile of $\Delta(T)$ 
as it approaches zero in a continuous way;
(ii) An ``upper'' $T_{c}^{+}$, corresponding to  the point $B$ in Fig.~\ref{Figure-2-SM}  and given by the maximum temperature at which the order parameter is nonzero.
In spite of this apparent uncertainty, here and in the Main Text we identify the values of the critical temperature $T_{c}$ throughout the BCS-BEC crossover as determined from the normal phase by the extended GMB approach of 
Ref.~\cite{PPPS-2018-Is}.
This choice rests on the fact that these theoretical values turn out to compare extremely well with the recent accurate measurements of $T_{c}$ made in Ref.~\cite{Koehl-2023s} across the BCS-BEC crossover through a pioneering application of machine learning methods.

Apart from the value of the critical temperature identified in this way, a certain degree of uncertainty on the temperature behavior of $\Delta(T)$ about $T_{c}$ should be conceded as again corresponding to a Ginzburg-like region,
similarly to what was pointed out in Ref.~\cite{Taylor-2009s}.
This is the additional reason why, in practice, we have avoided attempting a comparison between experiment and theory for values of physical quantities (like the zero-bias conductance (ZBC) reported in Figs.~3 and 4 of the Main Text) for temperatures
larger, say, than $0.85 T_{c}$, whereby the value of the gap parameter is identified by the point $C$ in Fig.~\ref{Figure-2-SM}.
This explains the choice of the temperature utilized in Fig.~4 of the Main Text.

\vspace{-0.2cm}
\section{RATIO $2\Delta/k_{B}T_{c}$ FOR  \\  QUASI-2D CONTACT POTENTIALS}
\label{sec:2D:contact_potential}

We provide here an argument for associating the coupling ratio $2\Delta/T_{c}$, as representative of the coupling strength for condensed-matter samples with quasi-2D character
(like films and layered materials, of interest for the experimental data on the tunneling spectra reported in the Main Text), with the coupling parameter $(k_{F} a_{F})^{-1}$ in terms of which theoretical calculations on the tunneling spectra
are available in 3D.

To this end, we discuss the questions of the correspondence 
(i) between the coupling ratio $2 \Delta/k_{B}T_{c}$ and the coupling parameter of a superfluid Fermi gas (separately in 3D and 2D) by relying on published calculations and 
(ii) between the coupling parameters of a superfluid Fermi gas in 3D and 2D.

\vspace{0.1cm}
\noindent
POINT (i):
We begin by relating the theoretical values obtained for $2\Delta/T_{c}$ in 3D and in the \emph{strictly\/}-2D case. 
This will be done by establishing a link between these two cases through a \emph{local\/} quantity like the Cooper pair size, which is expected not to be too much affected by dimensionality.
Lacking more detailed information, we shall then attribute to the experimental \emph{quasi\/}-2D samples an \emph{effective intermediate dimensionality\/} between 3D and 2D, 
by taking the arithmetic mean between the values of $2\Delta/T_{c}$ obtained in 3D and in the strictly-2D case.
This procedure will result in repositioning back toward the BCS regime condensed-matter samples which, based on the experimental values of the parameter $2\Delta/T_{c}$, were instead located in Ref.~\cite{Harrison_Chan-2022s} 
on the BEC side of unitarity.

\begin{figure}[t]
\begin{center}
\includegraphics[width=8.8cm,angle=0]{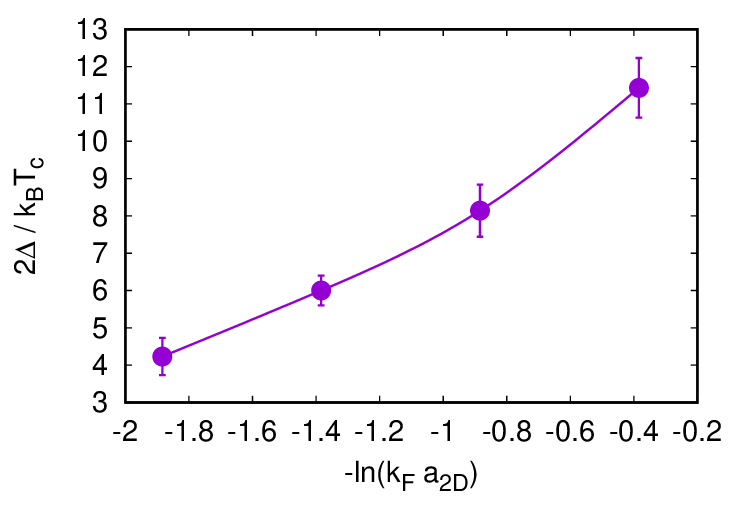}
\caption{Coupling ratio $2\Delta/k_{B}T_{c}$ as a function of the dimensionless coupling strength $-\ln(k_F a_{\rm 2D})$ for a 2D contact potential. 
              The ratio is constructed by using data for the zero temperature gap $\Delta$ from Ref.~\cite{Vitali-2017s} and data for the (BKT) critical temperature $T_{c}$ from Ref.~\cite{He-2022s}.}
\label{Figure-3-SM}
\end{center}
\end{figure}

The dependence of the coupling ratio $2\Delta/k_{B}T_{c}$ on the coupling strength for a contact potential in the strictly-2D case can be inferred from the Auxiliary-Field Quantum Monte Carlo (AFQMC) results, namely, 
from Ref.~\cite{Vitali-2017s} for the gap $\Delta$ at zero temperature and from Ref.~\cite{He-2022s} for the (BKT) critical temperature $T_{c}$. 
These results are utilized in  Fig.~\ref{Figure-3-SM}, to obtain the ratio $2\Delta/k_{B}T_c$ as a function of the 2D dimensionless coupling strength 
$\ln[(k_F a_{\rm 2D})^{-1}]=-\ln(k_F a_{\rm 2D})$. 
[For the weakest coupling point, for which Ref.~\cite{Vitali-2017s} does not report the AFQMC result for $\Delta$, we have used the GMB result provided in the same reference.] 
Similarly to Refs.~\cite{Feld-2011s,Pietila-2012s,Marsiglio-2015s}, the 2D scattering length $a_{\rm 2D}$ is here defined such that the binding energy $\epsilon_{0}=1/m a_{\rm 2D}^2$ of the two-body problem is formally 
given by the same expression of the 3D case. 
(This differs from the definition of the scattering length $a^*_{\rm 2D}$ used in Refs.~\cite{Vitali-2017s,He-2022s} by a constant factor, namely, $a_{\rm 2D} =a^*_{\rm 2D} e^\gamma/2$, $\gamma \simeq 0.577$ 
being the Euler constant.)
One sees from  Fig.~\ref{Figure-3-SM}  that the ratio $2\Delta/T_c$ increases monotonically with the 2D coupling strength.
This is similar to what is found in 3D, as shown Fig.~9 of Ref.~\cite{Deutsher-2019s} where the dependence of $2\Delta/k_{B}T_{c}$ on the coupling  $(k_{F} a_{F})^{-1}$ for the 3D contact potential is reported using the values of 
$\Delta$ and $T_{c}$ obtained by the extended GMB approach in Refs.~\cite{PPS-2018-IIs} and  \cite{PPPS-2018-Is}, respectively.

\begin{figure}[t]
\begin{center}
\includegraphics[width=8.8cm,angle=0]{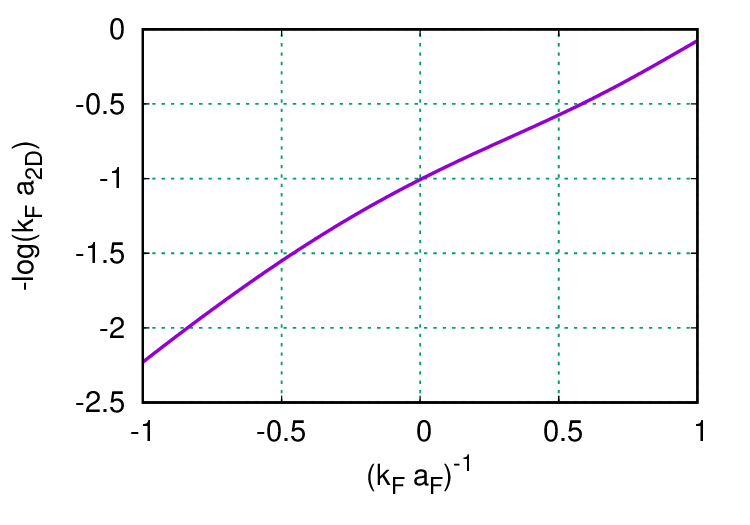}
\caption{Correspondence between 3D and 2D couplings, obtained from the equivalence of the dimensionless parameter $k_{F} \xi_{\rm pair}$ associated with the Cooper pair size.}
\label{Figure-4-SM}
\end{center}
\end{figure}

\vspace{0.1cm}
\noindent
POINT (ii): To make the above comparison meaningful, a criterion is needed that establishes a correspondence between the 3D and the strictly-2D coupling strengths. 
Following Ref.~\cite{PS-1994s}, we establish this correspondence in terms of the dimensionless quantity $k_{F}\xi_{\rm pair}$, where $\xi_{\rm pair}$ is the intra-pair coherence length (or Cooper pair size) obtained at $T = 0$ 
at the mean-field level \cite{Randeria-1990s,PS-1994s,Marini-1998s}.
This procedure was later validated experimentally in Ref.~\cite{Moritz-2022s}.
 Figure~\ref{Figure-4-SM}  reports this mapping between the 3D coupling $(k_{F} a_{F})^{-1}$ and the strictly-2D coupling $-\log(k_{F} a_{\rm 2D})$, which is obtained by finding, for given value 
of $(k_{F} a_{F})^{-1}$, the coupling $-\log(k_F a_{\rm 2D})$ such that the quantity $k_{F}\xi_{\rm pair}$ is the same in 2D and in 3D \cite{footnote-mapping}.

\begin{figure}[t]
\begin{center}
\includegraphics[width=8.8cm,angle=0]{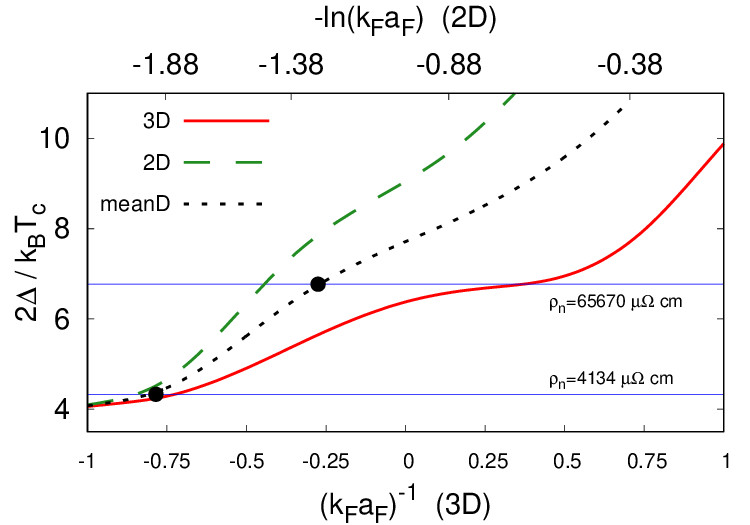}
\caption{Coupling ratio $2\Delta/k_{B}T_{c}$ vs the coupling strength for a 3D system (solid line and lower horizontal axis) and for a strictly-2D system (dashed line and upper horizontal axis).
              The correspondence between the upper and lower horizontal axes is provided by the mapping shown in  Fig.~\ref{Figure-4-SM}.
              The solid line is computed within the extended GMB approach of Ref.~\cite{PPPS-2018-Is}, while the dashed line is obtained by combining the QMC simulations of Refs.~\cite{Vitali-2017s} and \cite{He-2022s}.
              The dotted line is obtained by the arithmetic mean of the 3D and strictly-2D coupling ratios $2\Delta/k_{B}T_{c}$ for given respective coupling strengths, and represents the coupling ratio associated with an 
              effective ``intermediate'' dimensionality between 3D and 2D.
              The blue horizontal lines indicate the experimental values of the coupling ratios for the two samples considered specifically in Figs.~2 and 3 of the Main Text.
              The filled circles associate these coupling ratios to nominal values of the coupling $(k_{F} a_{F})^{-1}$, to be read from the lower horizontal axis.}
\label{Figure-5-SM}
\end{center}
\end{figure}

A full plot of the coupling ratio $2\Delta/k_{B}T_{c}$ as a function of both 3D and strictly-2D coupling strengths is shown in  Fig.~\ref{Figure-5-SM}.
In addition,  the arithmetic mean value  between the 3D and strictly-2D coupling strengths  is also reported in this figure.
 This is because the layered materials considered experimentally in our work are ``quasi'' two-dimensional superconductors, such that the question arises about their ``effective'' dimensionality being somewhat intermediate between 3D and 2D.
Lacking a full theory to account for intermediate dimensionalities in superfluid Fermi gases, we dwell on past experience in layered anti-ferromagnets (briefly recalled in Sec.~\ref{sec:anti_ferromagnet} below), 
according to which a rather small interlayer coupling proved sufficient to exhaust the crossover from 2D to 3D.
The mean value reported in  Fig.~\ref{Figure-5-SM}  then corresponds to an ``effective'' dimensionality between 3D and 2D, which is meant to take into account in an approximate way the actual ``intermediate'' dimensionality of the experimental samples. In practice, this procedure provides us with corresponding \emph{nominal\/} values of the coupling $(k_{F} a_{F})^{-1}$, which we have utilized in the numerical 3D calculations to obtain the single-particle density of states of the superfluid system and thus the differential conductance and the ensuing zero bias conductance, that are compared with the available experimental data in the Main Text.
In Fig.~\ref{Figure-5-SM}  this procedure is explicitly applied to the two grAl samples considered in Figs.~2, 3, and 4 of the Main Text, yielding the values $(k_{F} a_{F})^{-1}=-0.75$ for the lower and $(k_{F} a_{F})^{-1}=-0.25$ for the higher resistivity samples considered therein.
In particular, to the right-most sample reported in Fig.~4 of the Main Text we associate in this way an effective coupling parameter $(k_{F}a_{F})^{-1}$ with value $8$,
which corresponds essentially to unitarity according to  Fig.~\ref{Figure-5-SM}  above.
This result strongly supports our claim that the experimental values of the coupling ratio, once appropriately calculated, do not exceed (about) $8$.

Note from Fig.~\ref{Figure-5-SM}  that, upon approaching the BCS regime, the 3D and 2D values of the coupling ratio $2\Delta/k_{B}T_{c}$ tend to come close to each other, 
making irrelevant the above issue about an effective dimensionality when the Cooper pair size is much larger than the average inter-particle spacing.
On the other hand, in the unitary regime when the Cooper pair size is comparable with the average inter-particle spacing, the 3D and 2D values of the coupling ratio $2\Delta/k_{B}T_{c}$ are seen to deviate appreciably from each other.
In this case, one should be concerned with an effective dimensionality to be associated with the quasi-2D system.
Note further from Fig.~\ref{Figure-5-SM}  that the issue of an effective dimensionality has a strong consequence on the value of the nominal coupling $(k_{F} a_{F})^{-1}$ to be associated with a given sample.
For instance, while to the higher resistivity sample one would associate a strictly 3D coupling of about $+0.4$, the corresponding nominal coupling is about $-0.3$.
This has the effect of shifting the relevant coupling value from the BEC to the BCS side of unitarity.

The same kind of reasoning can as well be applied to the collection of data reported in Fig.~2(a) of Ref.~\cite{Harrison_Chan-2022s}, whereby, according to our procedure, the positions of all samples should somewhat be shifted toward
the BCS side.
This shift toward the BCS side adds to that pointed out in the Main Text, when the coherence energy scale $\Delta_{\mathrm{c}}$ is correctly used in the coupling ratio $2 \Delta / k_{B} T_{c}$ instead of the excitation gap $\Delta_{\mathrm{p}}$.
In particular, according to Fig.~\ref{Figure-5-SM}  above, samples with $2\Delta/T_{c} \approx 6.5$ in Fig.~2(a) of Ref.~\cite{Harrison_Chan-2022s} should not be associated with unitarity,
but rather with a nominal coupling of about $-0.4$.
It is interesting to mention that this value is just where there occurs the largest increase of the pseudo-gap energy scale due to pairing fluctuations  and of the critical temperature due to the combined effects of interaction and disorder \cite{PS-2013s}.
Once again, this shift toward the BCS side is consistent with one's expectation that for condensed-matter samples the Coulomb repulsion between electrons participating in pairing will eventually hamper the formation of truly composite bosons, that can instead form with neutral ultra-cold Fermi atoms in the BEC side of the crossover. 

\section{Analogy with the dimensional crossover between 3D and quasi-2D in a layered anti-ferromagnet}
\label{sec:anti_ferromagnet}

To reinforce the validity of the procedure described in Sec.~\ref{sec:2D:contact_potential}, an analogy can be drawn with the dimensional crossover occurring in a layered anti-ferromagnet, and, in particular, with the similarity between 3D 
and quasi-2D results obtained for local properties.

This system was studied theoretically both in the broken-symmetry phase below the critical (N\'{e}el) temperature $T_{N}$ \cite{MSS-1992s} and in the paramagnetic phase above $T_{N}$ \cite{MSS-1993s}, 
by varying the anisotropy ratio $\epsilon = J_{\mathrm{\perp}}/J_{\mathrm{\parallel}}$ between the inter-layer ($J_{\mathrm{\perp}}$) and intra-layer ($J_{\mathrm{\parallel}}$) nearest-neighbor couplings.
In that study, it was found that the parallel and perpendicular (instantaneous) spin-spin correlation functions both survive above $T_{N}$ for short spatial distances and that close to $T_{N}$ their magnitudes are comparable
even when $\epsilon$ is as small as $10^{-3}$.
This finding has led to the conclusion that short-range properties are not too much influenced by strong spatial anisotropy.
In addition, as the spins remain organized over moderate distances even above $T_{N}$, spin-wave excitations can still propagate over these distances as seen experimentally \cite{Mook-1973s}, although with a broadened frequency spectrum.

Correspondingly, the local spatial order in a superfluid Fermi system, which is preserved by the persistence of local pairing fluctuations above $T_{c}$, makes the single-particle excitation spectrum to resemble that below $T_{c}$, 
although with a frequency broadening due to the temporal decay of these local excitations that cannot propagate over long distances in the absence of (off-diagonal) long-range order \cite{PPSC-2002s}. 

\vspace{0.3cm}
\begin{figure}[h]
\begin{center}
\includegraphics[width=8.7cm,angle=0]{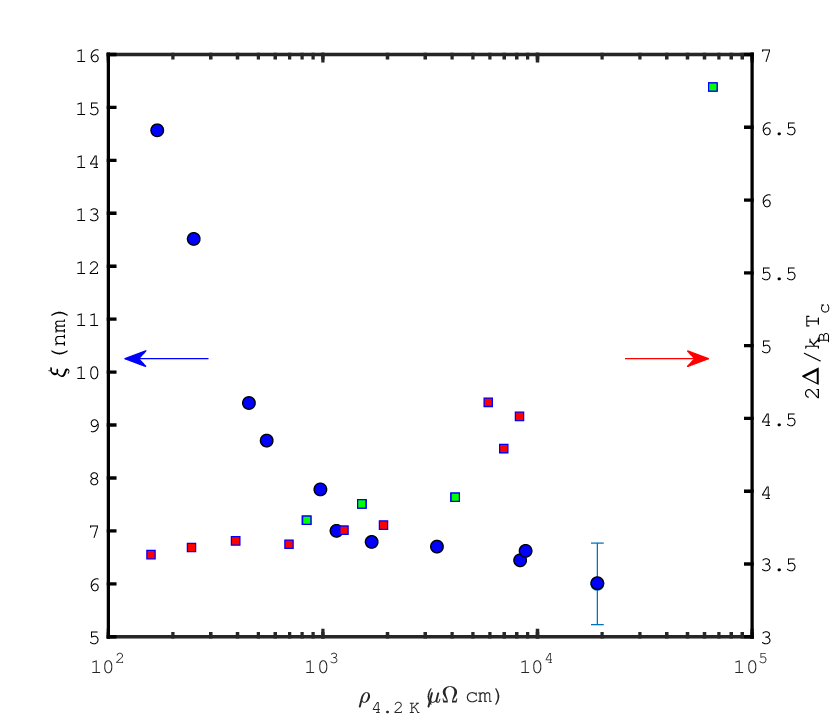}
\caption{The coherence length $\xi$ (left axis, filled dots) as obtained in previous work from upper critical field measurements \cite{MoshePRR-2020s,MoshePRB-2021s} 
              and the coupling ratio $2 \Delta/k_{B}T_{c}$ (right axis, filled squares) are shown as functions of the resistivity of the grAl samples analyzed in the present work. 
              The data point near 20,000 $\mu\Omega\ cm$ was obtained by linear extrapolation of the Maki parameter $\alpha$ for a sample for which the critical field was measured 
               at $T/T_{c} = 0.22$ (sample no.~9 in Ref.~\cite{MoshePRR-2020s}), such that the orbital field could be retrieved.}
\label{Figure-6-SM}
\end{center}
\end{figure}

\section{Coherence length}
\label{sec:coherence_length}

It was predicted in Ref.~\cite{PS-1996s} that the tendency toward a minimum in the coupling dependence of the low-temperature coherence length $\xi$ is a characteristic fingerprint of the approach 
to the unitarity regime within the BCS-BEC crossover.

This tendency is clearly consistent with the experimental data for $\xi$ reported in  Fig.~\ref{Figure-6-SM}  for the grAl samples, where highlighted in green are the four samples studied in the present work.
The figure also reports the values of the coupling ratio $2 \Delta/k_{B}T_{c}$ assigned to these samples.
It can be seen from this figure that, when the coherence length $\xi$ drops below a value of about 6–7 nm, the coupling ratio $2 \Delta/k_{B}T_{c}$ starts to increase significantly, in agreement with our expectation about the
approach to the unitary regime.

Recently, a similar behavior for $\xi$ was observed experimentally, in Ref.~\cite{Moritz-2022s} where the coherence length of an ultra-cold 2D Fermi gas (cf. Fig.~3a therein) is seen
to flatten out upon reaching unitarity from the BCS regime similarly to what would be found in 3D, as well as in Ref.~\cite{Jarillo-Herrero-2021s} where the Ginzburg–Landau coherence length (cf. Figs.~3b and 3c therein) 
is seen to approach the inter-particle distance around the optimal point in the phase diagram where $T_{c}$ is the highest.


\begin{center}
{\bf ACKNOWLEDGMENTS}
\end{center}

G.~C.~S. is grateful to A. Varlamov for discussions on the role of the AL and MT diagrammatic contributions for superconductors in the critical region above $T_{c}$.



\begin{thebibliography}{99}

\bibitem{Kanigel-2012} Y. Lubashevsky, E. Lahoud, K. Chashka, D. Podolsky, and A. Kanigel, \emph{Shallow pockets and very strong coupling superconductivity in} $FeSe_{x}Te_{1-x}$, Nat. Phys. {\bf 8}, 309 (2012).

\bibitem{Jarillo-Herrero-2021} J. M. Park, Y. Cao, K. Watanabe, T. Taniguchi, and P. Jarillo-Herrero, \emph{Tunable strongly coupled superconductivity in magic-angle twisted trilayer graphene\/}, Nature {\bf 590}, 249 (2021).

\bibitem{Deutsher-2021} A. G. Moshe, G. Tuvia, S. Avraham, E. Farber, and G. Deutscher, \emph{Tunneling study in granular aluminum near the Mott metal-to-insulator transition}\/, 
                                        Phys. Rev. B {\bf 104}, 054508 (2021).

\bibitem{Harrison_Chan-2022} N. Harrison and M. K. Chan, \emph{Magic gap ratio for optimally robust fermionic condensation and its implications for high-$T_{c}$ superconductivity\/},  
                                                 Phys. Rev. Lett. {\bf 129}, 017001 (2022).

\bibitem{Deutscher-1999} G. Deutscher, \emph{Coherence and single-particle excitations in the high-temperature superconductors\/}, Nature {\bf 397}, 410 (1999).

\bibitem{Levin-RMP-2024} Q. Chen, Z. Wang, R. Boyack, and K. Levin, \emph{When superconductivity crosses over from BCS to BEC\/}, Rev. Mod. Phys. {\bf 96}, 025002 (2024).


\bibitem{Physics-Reports-2018} G. Calvanese Strinati, P. Pieri, G. R\"{o}pke, P. Schuck, and M. Urban, \emph{The BCS-BEC crossover: From ultra-cold Fermi gases to nuclear systems\/}, Phys. Rep. {\bf 738}, 1 (2018).


\bibitem{Gonnelli-2002} R. S. Gonnelli, A. Calzolari, D. Daghero, L. Natale, G. A. Ummarino, V. A. Stepanov, and M. Ferretti, \emph{Doping dependence of the superconducting gap by Andreev reflection in 
                                       $\mathrm{Au}/\mathrm{La}_{2-x}\mathrm{Sr}_{x}\mathrm{CuO}_{4}$ point-contact junctions\/}, J. Phys. Chem. Solids {\bf 63}, 2369 (2002).


\bibitem{footnote-SM} See Supplemental Material for details on: i) experimental methods; ii) the relation between the coupling parameter $(k_{F} a_{F})^{-1}$ and the coupling ratio $2 \Delta/k_{B}T_{c}$; iii) the normalization procedure of the DOS spectra; iv) the emergence of a critical region upon approaching $T_c$; v) the issue of a meaningful comparison between the 3D and 2D values of the coupling ratio $2\Delta/k_{B}T_{c}$; vi) an analogy with the dimensional crossover occurring in a layered anti-ferromagnet; vii)  the coupling-dependence of the low-temperature coherence length. The Supplemental Material also includes Refs.~\cite{Deutsher-2021,Physics-Reports-2018, Moshe-2019,Harrison_Chan-2022,Fischer-1995,Fischer-2007,Kim-2022,PPS-2004,PPPS-2018-I,PPS-2018-II,Mahan-2000,LV-2009,Varlamov-1993,Haussmann-2007,Taylor-2009,Koehl-2023,Vitali-2017,He-2022,Feld-2011,Pietila-2012,Marsiglio-2015,PS-1994,Randeria-1990,Marini-1998,Moritz-2022-III,PS-2013,MSS-1992,MSS-1993,Mook-1973,PPSC-2002,PS-1996,MoshePRR-2020,Jarillo-Herrero-2021}.

\bibitem{Moshe-2019}
A.~G.~Moshe, E. Farber, and G. Deutscher, \emph{Optical conductivity of granular aluminum films near the Mott metal-to-insulator transition\/}, Phys. Rev. B {\bf 99}, 224503 (2019).

\bibitem{Fischer-1995} Ch. Renner and $\O$. Fischer, \emph{Vacuum tunneling spectroscopy and asymmetric density of states of $Bi_{2}Sr_{2}CaCu_{2}O_{8+\delta}$\/}, Phys. Rev. {\bf 51}, 9208 (1995).    

\bibitem{Fischer-2007} $\O$. Fischer, M. Kugler, I. Maggio-Aprile, and Ch. Berthod, \emph{Scanning tunneling spectroscopy of high-temperature superconductors\/}, Rev. Mod. Phys. {\bf 79}, 353 (2007).       

\bibitem{Kim-2022} H. Kim, Y. Choi, C. Lewandowski, A. Thomson, Y. Zhang, R. Polski, K. Watanabe, T. Taniguchi, J. Alicea, and S. Nadj-Perge, \emph{Evidence for unconventional superconductivity
                                 in twisted trilayer graphene\/}, Nature {\bf 606}, 494 (2022).   
                                 
\bibitem{PPS-2004} P. Pieri, L. Pisani, and G. Calvanese Strinati, \emph{BCS-BEC crossover at finite temperature in the broken-symmetry phase\/}, Phys. Rev. B {\bf 70}, 094508 (2004).                                                                                                                                                                                                                                          
  
 \bibitem{PPPS-2018-I} L. Pisani, A. Perali, P. Pieri, and G. Calvanese Strinati, \emph{Entanglement between pairing and screening in the Gorkov-Melik-Barkhudarov correction to the critical temperature 
                                   throughout the BCS-BEC crossover\/}, Phys. Rev. B {\bf 97}, 014528 (2018).

\bibitem{PPS-2018-II} L. Pisani, P. Pieri, and G. Calvanese Strinati, \emph{Gap equation with pairing correlations beyond the mean-field approximation and its equivalence to a Hugenholtz-Pines condition 
                                    for fermion pairs}, Phys. Rev. B {\bf 98}, 104507 (2018).                                              
\bibitem{Mahan-2000} G. D. Mahan, \emph{Many-Particle Physics\/} (Kluger, New York, 2000).       

\bibitem{LV-2009} A. Larkin and A. Varlamov, \emph{Theory of Fluctuations in Superconductors\/} (Oxford University Press, Oxford, 2009).

\bibitem{Varlamov-1993} V. V. Dorin, R. A. Klemm, A. A. Varlamov, A. I. Buzdin, and D. V. Livanov, \emph{Fluctuation conductivity of layered superconductors in a perpendicular magnetic field\/}, Phys. Rev. B {\bf 48}, 12951 (1993).

\bibitem{Haussmann-2007} R. Haussmann, W. Rantner, S. Cerrito, and W. Zwerger, \emph{Thermodynamics of the BCS-BEC crossover\/}, Phys. Rev. A {\bf 75}, 023610 (2007).     

\bibitem{Taylor-2009} E. Taylor, \emph{Critical behavior in trapped strongly interacting Fermi gases\/}, Phys. Rev. A {\bf 80}, 023612 (2009).                                                                                                                                                                                                                                                     
                                    
\bibitem{Koehl-2023} M. Link, K. Gao, A. Kell, M. Breyer, D. Eberz, B. Rauf, and M. K\"ohl, \emph{Machine learning the phase diagram of a strongly interacting Fermi gas\/}, Phys. Rev. Lett. {\bf 130}, 203401 (2023).       
                                                                                                                                                                                                                                                                                                  
\bibitem{Vitali-2017} E. Vitali, H. Shi, M. Qin, and S. Zhang, \emph{Visualizing the BEC-BCS crossover in a two-dimensional Fermi gas: Pairing gaps and dynamical response functions from \emph{ab initio\/} computations\/}, 
                                 Phys. Rev. A {\bf 96}, 061601 (2017).                                                                                                                                                                                                                                              
                                
\bibitem{He-2022} Y.-Y. He, H. Shi, and S. Zhang, \emph{Precision many-body study of the Berezinskii-Kosterlitz-Thouless transition and temperature-dependent properties in the two-dimensional Fermi gas\/}, 
                               Phys. Rev. Lett. {\bf 129}, 076403 (2022).                                                                                                                                                                                                                                                    
                               
\bibitem{Feld-2011} M. Feld, B. Fr\"{o}hlich, E. Vogt, M. Koschorreck, and M. K\"{o}hl, \emph{Observation of a pairing pseudogap in a two-dimensional Fermi gas\/}, Nature {\bf 480}, 75 (2011).                             

\bibitem{Pietila-2012} V. Pietil\"{a}, \emph{Pairing and radio-frequency spectroscopy in two-dimensional Fermi gases\/}, Phys. Rev. A {\bf 86}, 023608 (2012).                                                                                    

\bibitem{Marsiglio-2015} F. Marsiglio, P. Pieri, A. Perali, F. Palestini, and G. Calvanese Strinati, \emph{Pairing effects in the normal phase of a two-dimensional Fermi gas\/}, Phys. Rev. B {\bf 91}, 054509 (2015). 

\bibitem{PS-1994} F. Pistolesi and G. Calvanese Strinati, \emph{Evolution from BCS superconductivity to Bose condensation: Role of the parameter $k_{F} \xi$\/}, Phys. Rev. B {\bf 49}, 6356 (1994).                      

\bibitem{Randeria-1990} M. Randeria, J.~M. Duan, and L.~Y. Shieh, \emph{Superconductivity in a two-dimensional Fermi gas: Evolution from Cooper pairing to Bose condensation\/}, Phys. Rev. B {\bf 41}, 327 (1990).  

\bibitem{Marini-1998} M. Marini, F. Pistolesi, and G. Calvanese Strinati, \emph{Evolution from BCS superconductivity to Bose condensation: Analytic results for the crossover in three dimensions\/},
                                  Eur. Phys. J. B {\bf 1}, 151 (1998).                                                                                                                                                                                                                                                      

\bibitem{Moritz-2022-III} L. Sobirey, H. Biss, N. Luick, M. Bohlen, H. Moritz, and T. Lompe, \emph{Observing the influence of reduced dimensionality on fermionic superfluids\/}, Phys. Rev. Lett. {\bf 129}, 083601 (2022).
                                  
\bibitem{PS-2013} F. Palestini and G. Calvanese Strinati, \emph{Systematic investigation of the effects of disorder at the lowest order throughout the BCS-BEC crossover\/}, Phys. Rev. B {\bf 88}, 174504 (2013).  

\bibitem{MSS-1992} N. Majlis, S. Selzer, and G. Calvanese Strinati, \emph{Dimensional crossover in the magnetic properties of highly anisotropic antiferromagnets\/}, Phys. Rev. B {\bf 45}, 7872 (1992).                  

\bibitem{MSS-1993} N. Majlis, S. Selzer, and G. Calvanese Strinati, \emph{Dimensional crossover in the magnetic properties of highly anisotropic antiferromagnets. II. Paramagnetic phase\/}, 
                                 Phys. Rev. B {\bf 48}, 957 (1993).                                                                                                                                                                                                                                                          

 \bibitem{Mook-1973} H. A. Mook, J. W. Lynn, and R. M. Nicklow, \emph{Temperature dependence of the magnetic excitations in Nickel\/}, Phys. Rev. Lett. {\bf 30}, 556 (1973).            
                                                  
 \bibitem{PPSC-2002} A. Perali, P. Pieri, G. Calvanese Strinati, and C. Castellani, \emph{Pseudogap and spectral function from superconducting fluctuations to the bosonic limit\/}, Phys. Rev. B {\bf 66}, 024510 (2002). 
 
\bibitem{PS-1996} F. Pistolesi and G. Calvanese Strinati, \emph{Evolution from BCS superconductivity to Bose condensation: Calculation of the zero-temperature phase coherence length\/}, 
                              Phys. Rev. B {\bf 53}, 15168 (1996).
 
\bibitem{MoshePRR-2020} A. G. Moshe, E. Farber, and G. Deutscher, \emph{From orbital to Pauli-limited critical fields in granular aluminum films\/}, Phys. Rev. Res. {\bf 2}, 043354 (2020).                                                                                                                                                                                                                                                                                                      
                                                                                                                                                                                                                                                                                 


\bibitem{Tinkham-1980} M. Tinkham, \emph{Introduction to Superconductivity\/} (R. E. Krieger, Malabar, 1980). 

\bibitem{GMB-1961} L. P. Gor'kov and T. M. Melik-Barkhudarov, \emph{Contribution to the theory of superfluidity in an imperfect Fermi gas\/}, Sov. Phys. JETP {\bf 13}, 1018 (1961) [Zh. Eksp. Teor. Fiz. {\bf 40}, 1452 (1961)].   


\bibitem{Moritz-2022-I} H. Biss, L. Sobirey, N. Luick, M. Bohlen, J. J. Kinnunen, G. M. Bruun, T. Lompe, and H. Moritz, \emph{Excitation spectrum and superfluid gap of an ultracold Fermi gas\/}, 
                                   Phys. Rev. Lett. {\bf 128}, 100401 (2022).  

\bibitem{footnote-analytic_continuation} Inclusion of the GMB correction does not modify the procedure of analytic continuation from imaginary Matsubara frequencies $i \Omega_{\nu}$ to real frequencies $\omega$, 
                                                                used in the $t$-matrix approach of Ref.~\cite{PPS-2004} to obtain the single-particle spectral function $A(\mathbf{k},\omega)$, because here like in Ref.~\cite{PPS-2018-II} 
                                                                the GMB correction is considered only in the static limit.   
                                                                
\bibitem{PPS-2019} M. Pini, P. Pieri, and G. Calvanese Strinati, \emph{Fermi gas throughout the BCS-BEC crossover: Comparative study of $t$-matrix approaches with various degrees of self-consistency\/},
                                 Phys. Rev. B {\bf 99}, 094502 (2019).
                                 
\bibitem{Altshuler-1980} 
B.~L.~Altshuler,  A.~G.~Aronov, and P.~A.~Lee, \emph{Interaction Effects in Disordered Fermi Systems in Two Dimensions\/}, Phys. Rev. Lett. {\bf 44}, 1288 (1980)

\bibitem{Krasnov-2009} V. M. Krasnov, \emph{Temperature dependence of the bulk energy gap in underdoped ${\text{Bi}}_{2}{\text{Sr}}_{2}{\text{CaCu}}_{2}{\text{O}}_{8+\ensuremath{\delta}}$: 
                                        Evidence for the mean-field superconducting transition\/}, Phys. Rev. B {\bf 79}, 214510 (2009).

\bibitem{Deutscher-1977} G. Deutscher and S. A. Dodds, \emph{Critical-field anisotropy and fluctuation conductivity in granular aluminum films\/}, Phys. Rev. B {\bf 16}, 3936 (1977).                                 

\bibitem{Witkowski-2020} K. K. Witkowski and T. K. Kope\'{c}, \emph{Dimensional crossover in the Bose–Einstein condensation confined to anisotropic three‑dimensional lattices\/}, 
                                         J. Low Temp. Phys. {\bf 201}, 340 (2020).
                                         
\bibitem{Qiu-2021} D. Qiu, C. Gong, S. Wang, M. Zhang, C. Yang, X. Wang, and J. Xiong, \emph{Recent advances in 2D superconductors\/}, Adv. Mater. {\bf 33}, 2006124 (2021).                  

\bibitem{Nozieres-Pines-1958} P. Nozi\`{e}res and D. Pines, \emph{Electron Interaction in solids. General formulation\/}, Phys. Rev. {\bf 109}, 741 (1958).

\bibitem{DeGennes-1999} P. G. De Gennes, \emph{Superconductivity of Metals and Alloys\/} (Westview Press, Boulder,1999), Chap.~4.

\bibitem{Deutscher-2006} G. Deutscher, \emph{New superconductors: From granular to high $T_{c}$\/} (World Scientific, Singapore, 2006), Chap.~4.

                                         
\end{thebibliography}

\begin{thebibliography}{99}


\bibitem{MoshePRB-2021s} A. G. Moshe, G. Tuvia, S. Avraham, E. Farber, and G. Deutscher, \emph{Tunneling study in granular aluminum near the Mott metal-to-insulator transition\/}, 
                                           Phys. Rev. B {\bf 104}, 054508 (2021).                                                                                                                                                                                                                             


\bibitem{Physics-Reports-2018s} G. Calvanese Strinati, P. Pieri, G. R\"{o}pke, P. Schuck, and M. Urban, \emph{The BCS-BEC crossover: From ultra-cold Fermi gases to nuclear systems\/}, Phys. Rep. {\bf 738}, 1 (2018).

\bibitem{Deutsher-2019s} A. G. Moshe, E. Farber, and G. Deutscher, \emph{Optical conductivity of granular aluminum films near the Mott metal-to-insulator transition\/}, Phys. Rev. B {\bf 99}, 224503 (2019).

\bibitem{Harrison_Chan-2022s} N. Harrison and M. K. Chan, \emph{Magic gap ratio for optimally robust fermionic condensation and its implications for high-$T_{c}$ superconductivity\/},           
                                                 Phys. Rev. Lett {\bf 129}, 017001 (2022).   
                                                 

\bibitem{Fischer-1995s} Ch. Renner and $\O$. Fischer, \emph{Vacuum tunneling spectroscopy and asymmetric density of states of $Bi_{2}Sr_{2}CaCu_{2}O_{8+\delta}$\/}, Phys. Rev. {\bf 51}, 9208 (1995).    

\bibitem{Fischer-2007s} $\O$. Fischer, M. Kugler, I. Maggio-Aprile, and Ch. Berthod, \emph{Scanning tunneling spectroscopy of high-temperature superconductors\/}, Rev. Mod. Phys. {\bf 79}, 353 (2007).       

\bibitem{TTG-2022s} H. Kim, Y. Choi, C. Lewandowski, A. Thomson, Y. Zhang, R. Polski, K. Watanabe, T. Taniguchi, J. Alicea, and S. Nadj-Perge, \emph{Evidence for unconventional superconductivity
                                 in twisted trilayer graphene\/}, Nature {\bf 606}, 494 (2022).                                                                                                                                                                                                          


\bibitem{PPS-2004s} P. Pieri, L. Pisani, and G. Calvanese Strinati, \emph{BCS-BEC crossover at finite temperature in the broken-symmetry phase\/}, Phys. Rev. B {\bf 70}, 094508 (2004).  

\bibitem{PPPS-2018-Is} L. Pisani, A. Perali, P. Pieri, and G. Calvanese Strinati, \emph{Entanglement between pairing and screening in the Gorkov-Melik-Barkhudarov correction to the critical temperature 
                                   throughout the BCS-BEC crossover\/}, Phys. Rev. B {\bf 97}, 014528 (2018).                                                                                                                                                                             

\bibitem{PPS-2018-IIs} L. Pisani, P. Pieri, and G. Calvanese Strinati, \emph{Gap equation with pairing correlations beyond the mean-field approximation and its equivalence to a Hugenholtz-Pines condition 
                                    for fermion pairs\/}, Phys. Rev. B {\bf 98}, 104507 (2018).  

\bibitem{Mahan-2000s} G. D. Mahan, \emph{Many-Particle Physics\/} (Kluger, New York, 2000).       

\bibitem{LV-2009s} A. Larkin and A. Varlamov, \emph{Theory of Fluctuations in Superconductors\/} (Oxford University Press, Oxford, 2009).

\bibitem{Varlamov-1993s} V. V. Dorin, R. A. Klemm, A. A. Varlamov, A. I. Buzdin, and D. V. Livanov, \emph{Fluctuation conductivity of layered superconductors in a perpendicular magnetic field\/}, Phys. Rev. B {\bf 48}, 12951 (1993).

\bibitem{Haussmann-2007s} R. Haussmann, W. Rantner, S. Cerrito, and W. Zwerger, \emph{Thermodynamics of the BCS-BEC crossover\/}, Phys. Rev. A {\bf 75}, 023610 (2007).      

\bibitem{Taylor-2009s} E. Taylor, \emph{Critical behavior in trapped strongly interacting Fermi gases\/}, Phys. Rev. A {\bf 80}, 023612 (2009).                                                                                                       
\bibitem{Koehl-2023s} M. Link, K. Gao, A. Kell, M. Breyer, D. Eberz, B. Rauf, and M. K\"ohl, \emph{Machine learning the phase diagram of a strongly interacting Fermi gas\/}, Phys. Rev. Lett. {\bf 130}, 203401 (2023).                                                                                                                

\bibitem{Vitali-2017s} E. Vitali, H. Shi, M. Qin, and S. Zhang, \emph{Visualizing the BEC-BCS crossover in a two-dimensional Fermi gas: Pairing gaps and dynamical response functions from \emph{ab initio\/} computations\/}, 
                                 Phys. Rev. A {\bf 96}, 061601 (2017).                                                                                                                                                                                  
\bibitem{He-2022s} Y.-Y. He, H. Shi, and S. Zhang, \emph{Precision many-body study of the Berezinskii-Kosterlitz-Thouless transition and temperature-dependent properties in the two-dimensional Fermi gas\/}, 
                               Phys. Rev. Lett. {\bf 129}, 076403 (2022).                                                                                                                                                                                                         
\bibitem{Feld-2011s} M. Feld, B. Fr\"{o}hlich, E. Vogt, M. Koschorreck, and M. K\"{o}hl, \emph{Observation of a pairing pseudogap in a two-dimensional Fermi gas\/}, Nature {\bf 480}, 75 (2011).                             

\bibitem{Pietila-2012s} V. Pietil\"{a}, \emph{Pairing and radio-frequency spectroscopy in two-dimensional Fermi gases\/}, Phys. Rev. A {\bf 86}, 023608 (2012).                                                                                    

\bibitem{Marsiglio-2015s} F. Marsiglio, P. Pieri, A. Perali, F. Palestini, and G. Calvanese Strinati, \emph{Pairing effects in the normal phase of a two-dimensional Fermi gas\/}, Phys. Rev. B {\bf 91}, 054509 (2015). 


\bibitem{PS-1994s} F. Pistolesi and G. Calvanese Strinati, \emph{Evolution from BCS superconductivity to Bose condensation: Role of the parameter $k_{F} \xi$\/}, Phys. Rev. B {\bf 49}, 6356 (1994).                      

\bibitem{Randeria-1990s} M. Randeria, J.~M. Duan, and L.~Y. Shieh, \emph{Superconductivity in a two-dimensional Fermi gas: Evolution from Cooper pairing to Bose condensation\/}, Phys. Rev. B {\bf 41}, 327 (1990).  

\bibitem{Marini-1998s} M. Marini, F. Pistolesi, and G. Calvanese Strinati, \emph{Evolution from BCS superconductivity to Bose condensation: Analytic results for the crossover in three dimensions\/},
                                  Eur. Phys. J. B {\bf 1}, 151 (1998).                                                                                                                                                                                                                                                 
\bibitem{Moritz-2022s} L. Sobirey, H. Biss, N. Luick, M. Bohlen, H. Moritz, and T. Lompe, \emph{Observing the influence of reduced dimensionality on fermionic superfluids\/}, Phys. Rev. Lett. {\bf 129}, 083601 (2022).
                                  
\bibitem{footnote-mapping} A similar mapping using instead $\xi_{\rm pair}/r_n$, where $r_n$ is the average inter-particle distance, was utilized in Ref.~\onlinecite{Marsiglio-2015s}. 
Here we prefer to utilize instead $k_F \xi_{\rm pair}$, also in the light of its use in the recent experimental work of Ref.~\cite{Moritz-2022s}.                                                                 

\bibitem{PS-2013s} F. Palestini and G. Calvanese Strinati, \emph{Systematic investigation of the effects of disorder at the lowest order throughout the BCS-BEC crossover\/}, Phys. Rev. B {\bf 88}, 174504 (2013).  


\bibitem{MSS-1992s} N. Majlis, S. Selzer, and G. Calvanese Strinati, \emph{Dimensional crossover in the magnetic properties of highly anisotropic antiferromagnets\/}, Phys. Rev. B {\bf 45}, 7872 (1992).                  

\bibitem{MSS-1993s} N. Majlis, S. Selzer, and G. Calvanese Strinati, \emph{Dimensional crossover in the magnetic properties of highly anisotropic antiferromagnets. II. Paramagnetic phase\/}, Phys. Rev. B {\bf 48}, 957 (1993).                                                                                                                                                                      
\bibitem{Mook-1973s} H. A. Mook, J. W. Lynn, and R. M. Nicklow, \emph{Temperature dependence of the magnetic excitations in Nickel\/}, Phys. Rev. Lett. {\bf 30}, 556 (1973).                                                             
 
\bibitem{PPSC-2002s} A. Perali, P. Pieri, G. Calvanese Strinati, and C. Castellani, \emph{Pseudogap and spectral function from superconducting fluctuations to the bosonic limit\/}, Phys. Rev. B {\bf 66}, 024510 (2002). 
 

\bibitem{PS-1996s} F. Pistolesi and G. Calvanese Strinati, \emph{Evolution from BCS superconductivity to Bose condensation: Calculation of the zero-temperature phase coherence length\/}, Phys. Rev. B {\bf 53}, 15168 (1996).
 
\bibitem{MoshePRR-2020s} A. G. Moshe, E. Farber, and G. Deutscher, \emph{From orbital to Pauli-limited critical fields in granular aluminum films\/}, Phys. Rev. Res. {\bf 2}, 043354 (2020).                                                                                                                                                                                                                 
\bibitem{Jarillo-Herrero-2021s} J. M. Park, Y. Cao, K. Watanabe, T. Taniguchi, and P. Jarillo-Herrero, \emph{Tunable strongly coupled superconductivity in magic-angle twisted trilayer graphene\/}, Nature {\bf 590}, 249 (2021).                                                                                                                                                                                                                                                                                 
\end{thebibliography}
\end{document}